\documentclass[aps,prfluids,onecolumn,a4]{revtex4}
\usepackage{epsfig}
\usepackage[utf8]{inputenc}
\usepackage[english]{babel}
\usepackage{setspace,amsmath}
\usepackage{comment}
\usepackage{graphicx}
\usepackage{amsmath}
\usepackage{amssymb}
\usepackage{color}
\usepackage{bm}

\newcommand{\ba}{\begin{eqnarray}}

\begin{document}

\title{Assessing non-Oberbeck-Boussinesq effects of convection in cryogenic helium}
\author{Michal Macek$^1$}
\author{Georgy Zinchenko$^2$}
\author{V{\v e}ra Musilov\'a$^1$}
\author{Pavel Urban$^1$}
\author{J\"org Schumacher$^{2,3}$}
 \affiliation{
 $^1$The Czech Academy of Sciences, Institute of Scientific Instruments, Kr\'alovopolsk\'a 147, CZ-61264 Brno, Czech Republic\\
 $^2$Institut für Thermo- und Fluiddynamik, Technische Universität Ilmenau, Postfach 100565, D-98684 Ilmenau, Germany\\
 $^3$Tandon School of Engineering, New York University, New York City, NY 11201, USA}

\date{\today}

\begin{abstract}
The present study investigates the non-Oberbeck-Boussinesq (NOB) effects which arise due to the temperature dependence of material properties in cryogenic helium experiments of turbulent Rayleigh-B\'{e}nard convection. They are manifest as a difference of the measured mean temperature at the center of the closed cell, $T_c$, from the arithmetic mean temperature obtained from the prescribed fixed and uniform temperatures at the top and bottom copper plates of the apparatus, $T_m=(T_{\rm bot}+T_{\rm top})/2$. Therefore, the material properties such as specific heat at constant pressure, dynamic viscosity, thermal conductivity, the isobaric expansivity, and the mass density are expanded into power series with respect to temperature up to the quadratic order with coefficients obtained from the software package HEPAK. A subsequent nonlinear regression that uses deep convolutional networks delivers a dependence of the strength of non-Oberbeck-Boussinesq effects in the pressure--temperature parameter plane. Strength of the NOB effects is evaluated via the deviation of the mean temperature profile $\xi_\mathrm{NOB} \equiv T_m - T_c$ from the top/bottom-symmetric Oberbeck-Boussinesq case $\xi_\mathrm{NOB}=0$. Training data for the regression task are obtained from 236 individual long-term laboratory measurements at different Rayleigh numbers which span 8 orders of magnitude.    
\end{abstract}

\maketitle

\section{Introduction}
Controlled laboratory experiments of turbulent Rayleigh-B\'enard convection (RBC) are one pillar of turbulence research to obtain a deeper understanding of the physical transfer processes and their coupling to statistical properties and structures, both in the bulk and the boundary layers of buoyancy-driven flows \cite{Kadanoff2001,Ahlers2009,Chilla2012,Verma2018}. The highest Rayleigh numbers $Ra$ for fluid flows at Prandtl numbers $Pr\simeq 1$ are obtained in two gases, either compressed sulphur hexafluoride, SF$_6$, \cite{He2012,He2012a} or cryogenic helium, $^4$He, the latter of which is cooled down to a few Kelvin \cite{Chavanne1997,Niemela2000,Niemela2003,Roche2010,Urban2011,Urban2012,Urban2019,Roche2020}. While the Rayleigh number $Ra$ quantifies the thermal driving of convective turbulence, the Prandtl number $Pr$ is the ratio of molecular momentum to temperature diffusion. Together with a third parameter, the aspect ratio of the exclusively used cylindrical closed vessels $\Gamma=D/H$ with cell diameter $D$ and cell height $H$, these three dimensionless numbers determine the control parameters of the experiments and are subsequently used to quantify the response of the apparatus in the form of power laws of the turbulent momentum and heat transfer. The latter are quantified by the dimensionless Reynolds and Nusselt numbers, $Re$ and $Nu$ \cite{Ahlers2009,Chilla2012}. The present study is focused to the experiments in cryogenic helium $^4$He. 

The Rayleigh-B\'{e}nard convection model incorporates the Oberbeck-Boussinesq (OB) approximation \cite{Chilla2012,Verma2018} which considers the working fluid as incompressible. In addition, the mass density field of the fluid, $\rho({\bm x},t)$ is taken as a linear function of the temperature field $T({\bm x},t)$ and given by
\begin{equation}
\rho({\bm x},t)=\rho_{\rm ref}[1-\alpha (T({\bm x},t)-T_{\rm ref})] \quad \mbox{with} \quad \alpha=-\frac{1}{\rho}\frac{\partial \rho}{\partial T}\Big|_p\,,
\label{Bouss1}
\end{equation}
with the isobaric thermal expansion coefficient or expansivity $\alpha$. This dependence is incorporated in the volume forces and thus couples the temperature field to the momentum balance. Quantities $\rho_{\rm ref}$ and $T_{\rm ref}$ are reference magnitudes of density and temperature, respectively. One important consequence of the OB approximation is that statistical properties, such as mean profiles, in the lower and upper halves of the convection cell including the corresponding viscous and thermal boundary layers, are symmetric with respect to the midplane at $z=H/2$. Consequently, 
\begin{equation}
T_c:=\Big\langle T\left(z=\frac{H}{2}\right)\Big\rangle=\frac{T_{\rm bot}+T_{\rm top}}{2}=:T_m\,,
\label{Bouss2}
\end{equation}
in laboratory experiments with the prescribed fixed and uniform temperatures at the top and bottom, $T_{\rm top}$ and $T_{\rm bot}$. 

\begin{figure}[t]
\centering
\includegraphics[width=0.54\linewidth]{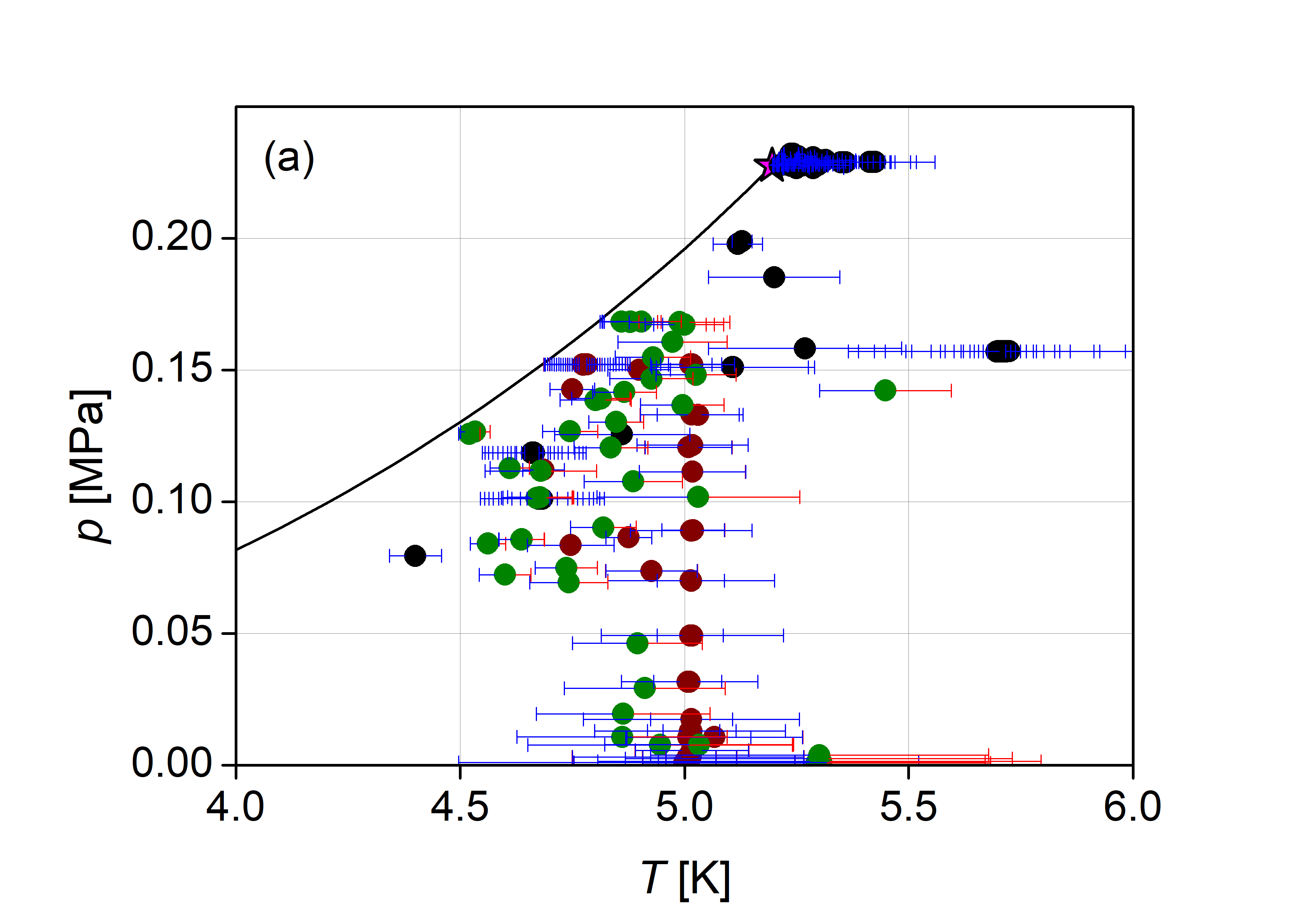}
\includegraphics[width=0.46\linewidth]{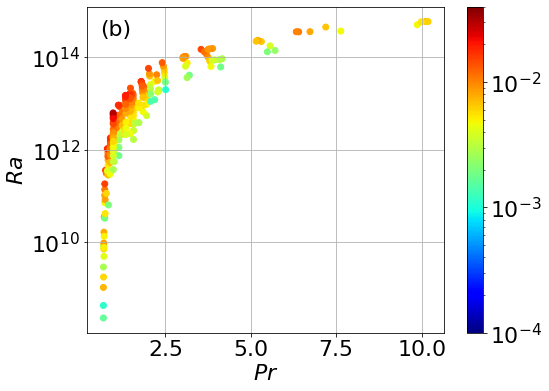}
\caption{Summary of the operating points $(p_m, T_m)$ of the cryogenic Rayleigh-B\'{e}nard experiments. (a) The mean temperatures $T_m$ and the temperature range $\Delta T=T_{\rm bot}-T_{\rm top}$ at a mean pressure $p_m$ are provided in the $p$--$T$ parameter plane. The 'errorbars' stemming from the $(p_m$,$T_m$) points denote the ranges of $\Delta T$ between the cold (blue) and hot (red) plate temperatures $T_t$, $T_b$, respectively. The solid line marks the saturated vapor curve (SVC) and the star symbol in the center of the figure indicates the critical point (CP) with $T_{\rm cri} = 5.195$~K, $p_{\rm cri} = 0.228$~MPa. The different colors of experimental $(p_m$,$T_m$) points correspond to new data (green), and data published in Refs.~\cite{Urban2014} (black) and~\cite{Urban2021} (brown). (b) Color-coded non-Boussinesq parameter $\xi_{\rm NOB}$ at the $Ra, Pr$ control parameters for the experiments shown in panel (a), given in units of Kelvin.}
\label{fig:PTexperiments}
\end{figure}

Cryogenic RBC experiments at the highest Rayleigh numbers have to be operated close to the critical point (CP) of He \cite{Niemela2000,Urban2011,Urban2012,Urban2019}. At this point, where the saturated vapor curve (SVC) representing the phase boundary between the gas and liquid state ends, the material properties of the working fluid such as  specific heat at constant pressure, $C_p$, dynamic viscosity $\mu$, or thermal conductivity $\lambda$ fluctuate strongly. This is considered as one possible source of the deviations from the Boussinesq limit, which are experimentally probed by a violation of \eqref{Bouss2}. In other words, non-Boussinesq (NOB) effects are detected as $T_c\ne T_m$. It is exactly this deviation which we want to explore in detail in the present work for cryogenic $^4$He. Therefore, we define the non-Oberbeck-Boussinesq parameter 
\begin{equation}
\xi_{\rm NOB}(p_m,T_m,\chi_k) :=T_m-T_c\,,
\label{ksiNOB}
\end{equation}
where $\chi_k$ are for now a short-hand notation for material properties which will be specified further below. Figure \ref{fig:PTexperiments} (panel a) summarizes the operating points within the $p$--$T$ diagram for cryogenic experiments conducted in the apparatus of the group in Brno (Czech Republic) \cite{Urban2010}. We indicate the mean temperature as well as the range of the applied outer temperature difference $\Delta T=T_{\rm bot}-T_{\rm top}>0$ at the mean pressure $p = p_m$. It is seen that a number of measurements are close to the phase boundary (solid black curve) and a few even in the vicinity of the critical point (magenta star). In panel (b), the measured values of $\xi_{\rm NOB}$ are plotted in the phase diagram of the two control parameters, $Ra$ and $Pr$. It is important to note that $Ra$ and $Pr$ are control parameters characterizing OB convection, thus ideally $\xi_{\rm NOB} = 0$ independent of $Ra$ and $Pr$. The experimentally observed values $\xi_{\rm NOB} \neq 0$  unambiguously indicate presence of NOB effects and must be captured introducing additional control parameters. We define and discuss a suitable set below.

In this work, we will systematically investigate the non-Boussinesq effects in cryogenic helium experiments at high Rayleigh numbers spanning a range of $10^7\lesssim Ra\lesssim 10^{15}$ performed in Brno. We quantify the deviation of the center temperature $T_c$ from $T_m$ in the $p$--$T$ diagram by means of nonlinear regression applying deep neural networks, i.e., determine $\xi_{\rm NOB}(p_m, T_m)$. This regression proceeds in three different levels of refinement, partly based on a perturbative expansion of the temperature dependence of essential material parameters and state variables $p$, $T$ as described in the next section. Furthermore, we aim to identify which of the variations of the material parameters are the most important ones for the magnitude of the NOB parameter. The complex material dependencies including discontinuities at the phase boundary and the singularities at the critical point are tabulated in the software package HEPAK written by V. Arp et al. \cite{Arp1998} and retailed by Horizon Technologies Inc \cite{Horizon}. They allow us to quantify the prefactors of the polynomial expansions of the material parameters and state variables at different orders. Starting point is the set of fully compressible equations of motion which was outlined by Gray and Giorgini~\cite{Gray1976}. Material parameter dependencies have been also systematically discussed for high-Rayleigh-number experiments in compressed SF$_6$ in refs. \cite{Shishkina2016,Weiss2018,Yik2021}.  

Our paper is organized as follows. Section~\ref{sec:ParEq} introduces the fully compressible equations of motion and discusses the resulting polynomial expansions for the material properties. In Sec. \ref{sec:properties}, we explore the basic state and transport properties of $^4$He as a function of pressure $p$ and temperature $T$ in connection with the ($p_m,T_m$) operating points of all realized RBC experiments within the respective $\Delta T$ region. Here, we also discuss the second mechanism to NOB convection, the compressibility effects which are shown to be negligible in the present setup. In Sec.~\ref{sec:Apparatus}, we briefly overview the essential features of the experimental set-up for cryogenic RBC.  Section~\ref{sec:ML} discusses the nonlinear regression results. In the last section, we give the conclusions and outlook. Technical details of the deep neural networks and an error analysis of the machine learning procedures are discussed in the appendices A and B.  

\section{Perturbative expansion of the equations of motion}\label{sec:ParEq}

Convective motions in a fluid layer are described by the set of three balance equations involving the continuity equation for the mass balance, the Navier-Stokes equations for the momentum balance, and the energy balance equation. Following here the textbook by Batchelor \cite{Batchelor1960} and the seminal work by Gray and Giorgini \cite{Gray1976}, they are given by
\begin{align}
\label{eq:Cont}
\frac{D \rho }{D t}&=-\rho \frac{\partial u_j}{\partial x_j}, \\ 
\label{eq:NS}
\rho \frac{D u_i}{D t}&=-\frac{\partial p}{\partial x_i}-\rho g_i \alpha T+\frac{\partial}{\partial x_j} (\mu \Gamma _{ij}), \\ 
\label{eq:Heat}
\rho C_p\frac{D T}{D t}-\alpha T\frac{Dp}{Dt}&=\frac{\partial }{\partial x_j}\left( \lambda \frac{\partial T}{\partial x_j} \right)+\mu \Phi,
\end{align}
where $D\bullet/Dt = \partial\bullet/\partial t + {\bm u}\cdot\nabla\bullet$ is the material derivative and 
\begin{equation}
	\Gamma _{ij}=\frac{\partial u_i}{\partial x_j}+\frac{\partial u_j}{\partial x_i}-\frac{2}{3}\frac{\partial u_k}{\partial x_k}\delta _{ij}
\end{equation}
is the rate of strain tensor in the compressible case. Furthermore,
\begin{equation}	
\Phi =\frac{1}{2}{\Gamma_{ij}}\left( \frac{\partial u_i}{\partial x_j}+\frac{\partial u_j}{\partial x_i} \right),
\end{equation}
is the dissipation function and $ g_i =\left(0,0,-g\right)$ the vector that contains the acceleration due to gravity $g$ in vertical direction. The bulk viscosity is set to zero.

A simplified form of the convection equations \eqref{eq:Cont}--\eqref{eq:Heat} is obtained in the form of the OB approximation, once the dynamical viscosity $\mu$ , the thermal conductivity $\lambda$, the isobaric thermal expansivity $\alpha$, and the specific heat $C_p$ at constant pressure are taken as constants. Furthermore, the mass density is taken to be constant $\rho = \rho_0$, such that the flow is basically incompressible, and heating due to pressure variations remains subdominant, which results to 
\begin{align}
\label{eq:Cont0}
\frac{\partial u_j}{\partial x_j}&=0, \\ 
\label{eq:NS0}
\rho_0 \frac{D u_i}{D t}&=-\frac{\partial p}{\partial x_i}-\rho_0 g_i \alpha T+\mu\frac{\partial^2 u_{i}}{\partial x_j^2}, \\ 
\label{eq:Heat0}
\rho_0 C_p\frac{D T}{D t}&=\lambda \frac{\partial^2 T }{\partial x_j^2}.
\end{align}
This implies that the speed of sound is much larger than the typical convection velocity, the free-fall velocity $U_{\rm f} = \sqrt{g\alpha\Delta T H}$ for small $\alpha\Delta T$. In all expressions above we have used the Einstein summation convention. 

Our objective is to evaluate the importance of individual terms of the compressible equations \eqref{eq:Cont}--\eqref{eq:Heat} beyond the OB limit systematically and analyze their effects on the mean temperature profiles in the thermal convection experiments in cryogenic helium. We will therefore assume that all material parameters are functions of the temperature only and that their pressure dependence is much less significant for the present experimental conditions. They can then be approximated by Taylor expansions with respect to $T$, which we will follow up to the quadratic expansion term. This results to the following expressions
\begin{align}
		\rho &={{\rho }_{0}}\left( 1-{{\alpha }_{0}}\delta T+{{\beta }_{0}}\delta {{T}^{2}} \right), \label{Mat1}
	\\ 
		\mu &={{\mu }_{0}}\left( 1+{{m}_{10}}\delta T+{{m}_{20}}\delta {{T}^{2}} \right), \label{Mat2}
	\\ 
		{{C}_{p}}&={{C}_{p0}}\left( 1+{{c}_{10}}\delta T+{{c}_{20}}\delta {{T}^{2}} \right), \label{Mat3}
	\\ 
		\lambda &={{\lambda }_{0}}\left( 1+{{l}_{10}}\delta T+{{l}_{20}}\delta {{T}^{2}} \right), \label{Mat4}		
	\\ 
		\alpha &={{\alpha }_{0}}\left( 1+{{a}_{10}}\delta T+{{a}_{20}}\delta {{T}^{2}} \right). \label{Mat5}
\end{align}
where $\delta T=T-T_m$. The first index in coefficients defines the term number in Taylor series decomposition,  index $0$ refers to the mean temperature $T_m$ and the minus in (\ref{Mat1}) follows a convention usual in fluid dynamics. All constants in the expansions \eqref{Mat1}--\eqref{Mat5} are determined from quadratic interpolations of the respective material properties at three temperature values $T_{\rm top}$, $T_m$, and  $T_{\rm bot}$ at a given pressure value $p$ using HEPAK \cite{Arp1998}. 

The substitution of Eqs.~\eqref{Mat1}--\eqref{Mat5} into the compressible fluid equations~\eqref{eq:Cont}--\eqref{eq:Heat} and the subsequent performance of a number of transformations leads to a set of a dimensionless convection equations which include the NOB effects up to second order with respect to temperature. They are given by
\begin{align}
\label{13}
-\varepsilon_1\frac{D\theta }{Dt}+2\varepsilon_2\theta \frac{D\theta }{Dt} &=-\left( 1-\varepsilon_1\theta +\varepsilon_2\theta^2  \right)\frac{\partial u_j}{\partial x_j},
\\ 
\label{14}
\left( 1-\varepsilon_1\theta+\varepsilon_2\theta^2 \right)\frac{Du_i}{D t} &=-\frac{\partial \left( p-p_s \right)}{\partial x_i}+\left[ \left( \theta -\theta_s \right)-\frac{\varepsilon_2}{\varepsilon_1}\left( \theta^2\!-\!\theta _s^2 \right) \right] k_i +\nonumber
\\ 
& \frac{1}{Re_{\rm f}}\left( 1+\varepsilon_3\theta +\varepsilon_4\theta^2 \right)\frac{\partial \Gamma _{ij}}{\partial x_j}+\frac{1}{Re_{\rm f}}\left(\varepsilon_3+2\varepsilon_4\theta  \right)\Gamma_{ij}\frac{\partial \theta }{\partial x_j},
\\ 
\label{15}
 \left( 1-\varepsilon_1\theta + \varepsilon_2\theta^2 \right)\left( 1 + \varepsilon_5\theta + \varepsilon_6\theta^2 \right)\frac{D\theta }{Dt}&=\frac{\tilde D}{Re_{\rm f}}\left( 1 + \varepsilon_3\theta + \varepsilon_4\theta^2 \right) \Phi  + \nonumber
\\ 
& \frac{1}{Re_{\rm f} Pr}\left( 1 + \varepsilon_7\theta + \varepsilon_8\theta^2 \right)\frac{\partial^{2}\theta}{\partial x_j^2} + \frac{1}{Re_{\rm f} Pr }\left(\varepsilon_7 + 2 \varepsilon_8\theta  \right)\left( \frac{\partial \theta}{\partial x_j} \right)^2 +\nonumber
\\
& \tilde{D}\left( 1 + \varepsilon_9\theta + \varepsilon_{10}\theta^2 \right)\left[\varepsilon_1\frac{D\left( p-p_s \right)}{Dt}-\left(1 - \varepsilon_1\theta_s+ \varepsilon_2\theta_s^2 \right)u_3 \right] (\theta + \Tilde{T}_m)\,.
\end{align} 
Here,
\begin{equation}
\theta=\frac{\delta T}{\Delta T}=\frac{T-T_m}{T_{\rm bot}-T_{\rm top}}\quad \mbox{and} \quad\Tilde{T}_m=\frac{T_m}{\Delta T}=\frac{T_m}{T_{\rm bot}-T_{\rm top}}\,,
\end{equation}
and $\theta_s$ and $p_s$ are temperature and pressure equilibrium (static heat conduction) profiles, respectively. Furthermore, we define in the energy balance a dimensionless parameter 
\begin{equation}\label{eq:TildeD}
\tilde{D}=\frac{g\alpha_0 H}{C_{p0}}\,,
\end{equation}
which is denoted as the dissipation number \cite{Verhoeven2015}. This number relates the dry adiabatic lapse rate $g/C_{p0}$ to the characteristic temperature drop $\alpha_0H$. The velocity was made non-dimensional by $U_{\rm f}$. The Prandtl number reads $Pr=\nu_{0}/\kappa_{0}$ with the temperature diffusivity $\kappa_0=\lambda_0/(\rho_0 C_{p0})$, the free-fall Reynolds number $Re_{\rm f}= U_{\rm f} H/\nu_{0}$, and the unit vector in the momentum balance points into the positive $z$--direction, $k_{i}=(0,0,1)$. The free-fall Reynolds number follows to $Re_{\rm f}=\sqrt{Ra/Pr}$ where $Ra$ is the Rayleigh number 
\begin{equation}
Ra=\frac{\alpha_0}{\nu_0\kappa_0} g \Delta T H^3\,.
\end{equation}

\begin{table}[h!]
\begin{tabular}{lll}
\hline\hline
Quantity  & First order expansion$\quad$ &  
            Second order expansion$\quad$\\
\hline
Mass density $\rho$ & ${{\varepsilon }_{1}}  ={{\alpha }_{0}}\Delta T$ & 
                       ${{\varepsilon }_{2}}  ={{\beta }_{0}}{{\Delta T}^{2}}$\\ 
Dynamic viscosity $\mu$ & ${{\varepsilon }_{3}} ={{m}_{10}}\Delta T$ &
                          ${{\varepsilon }_{4}}  ={{m}_{20}}{{\Delta T}^{2}}$ \\ 
Specific heat $C_p$ & ${{\varepsilon }_{5}} ={{c}_{10}}\Delta T$ &
                      ${{\varepsilon }_{6}}  ={{c}_{20}}{{\Delta T}^{2}}$\\
Thermal conductivity $\lambda$ & ${{\varepsilon }_{7}} ={{l}_{10}}\Delta T$ &
                     ${{\varepsilon }_{8}} ={{l}_{20}}{{\Delta T}^{2}}$ \\
Isobaric expansion coefficient $\alpha\quad$ & ${{\varepsilon }_{9}} ={{a}_{10}}\Delta T$ & ${{\varepsilon }_{10}} ={{a}_{20}}{{\Delta T}^{2}}$ \\
\hline\hline
\end{tabular}
\caption{List of all Taylor expansion parameters $\varepsilon_i$ for $i=1,...,10$.}
\label{Tab_eps}
\end{table}
Finally, in table \ref{Tab_eps}, we list all parameters $\varepsilon_i$ which were used in Eqs. \eqref{13}--\eqref{15} for first and second order expansions. Setting all these expansion parameters $\varepsilon_i$ and $\tilde{D}$ to zero recovers the OB equations \eqref{eq:Cont0}--\eqref{eq:Heat0}~\cite{Ahlers2009,Chilla2012}. Expansion parameters with an odd-number index are for the linear expansion while those with an even index are for the quadratic order. Notice that $\tilde{D}$, similar to the OB control parameters $Ra$ and $Pr$ depends only on local values of the material properties at the reference temperature $T_m$, and---like the Prandtl number $Pr$---is independent of $\Delta T$. In this sense it differs from $\varepsilon_i$, which are ``non-local'' and depend on the temperature derivatives at $T_m$ as well as $\Delta T$. The regression algorithms in Sec. \ref{sec:ML} will proceed in incremental steps, i.e., consider linear expansions only at first and incorporate second order subsequently. 

\section {State and transport properties of cryogenic $^4$He}
\label{sec:properties}

\begin{figure*}[t]
\centering
\includegraphics[width=0.45\linewidth]{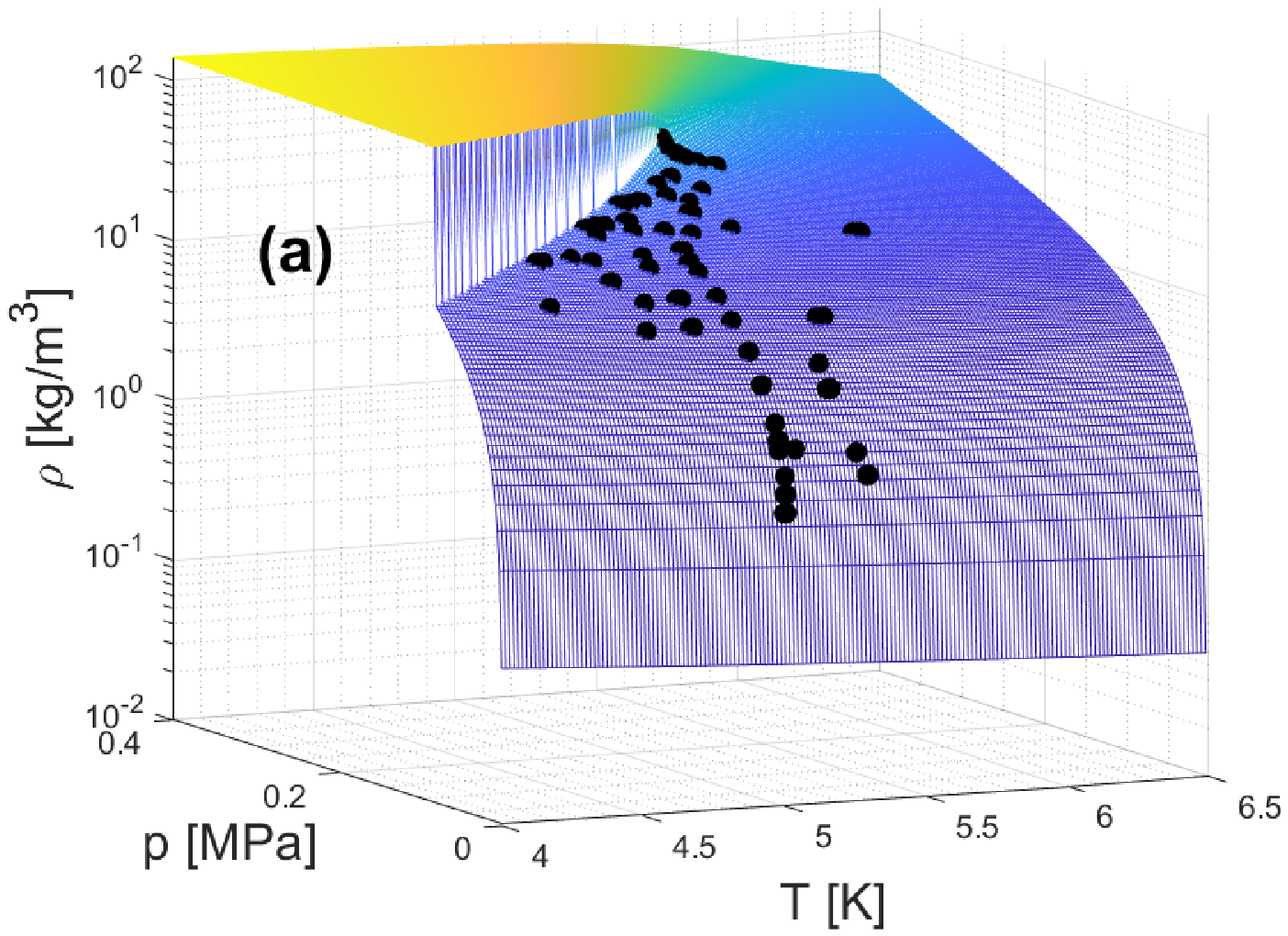}
\includegraphics[width=0.45\linewidth]{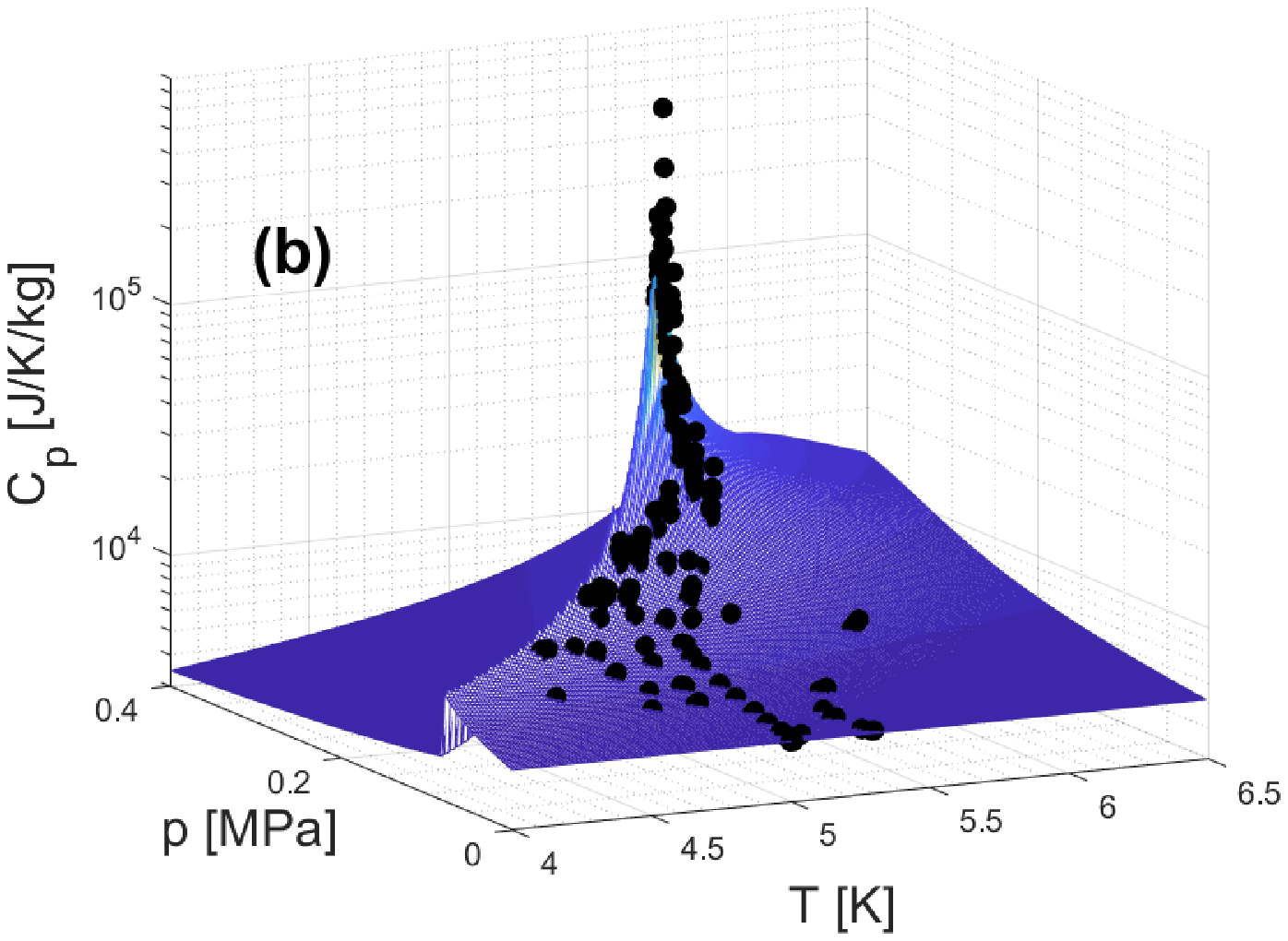}
\includegraphics[width=0.45\linewidth]{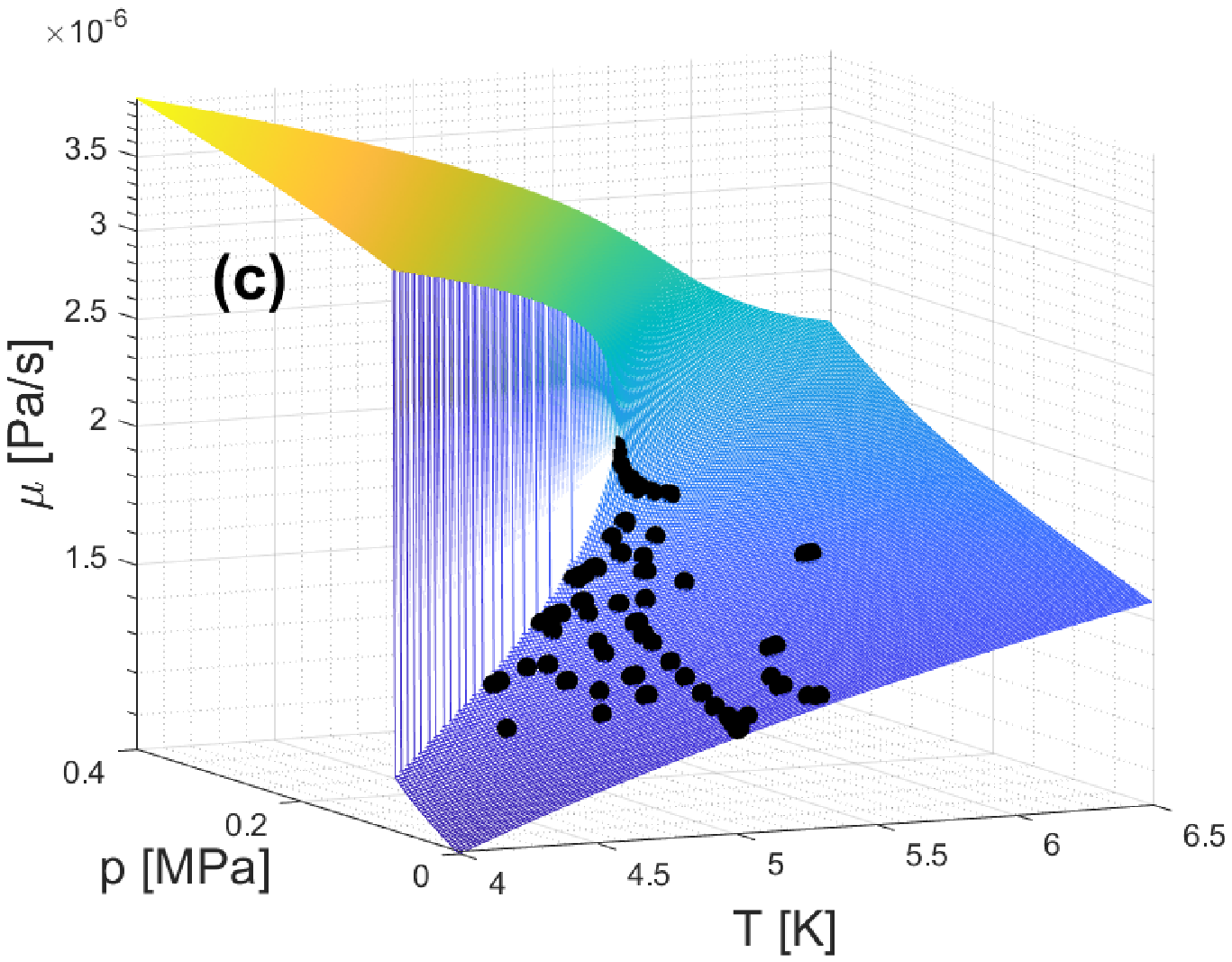}
\includegraphics[width=0.45\linewidth]{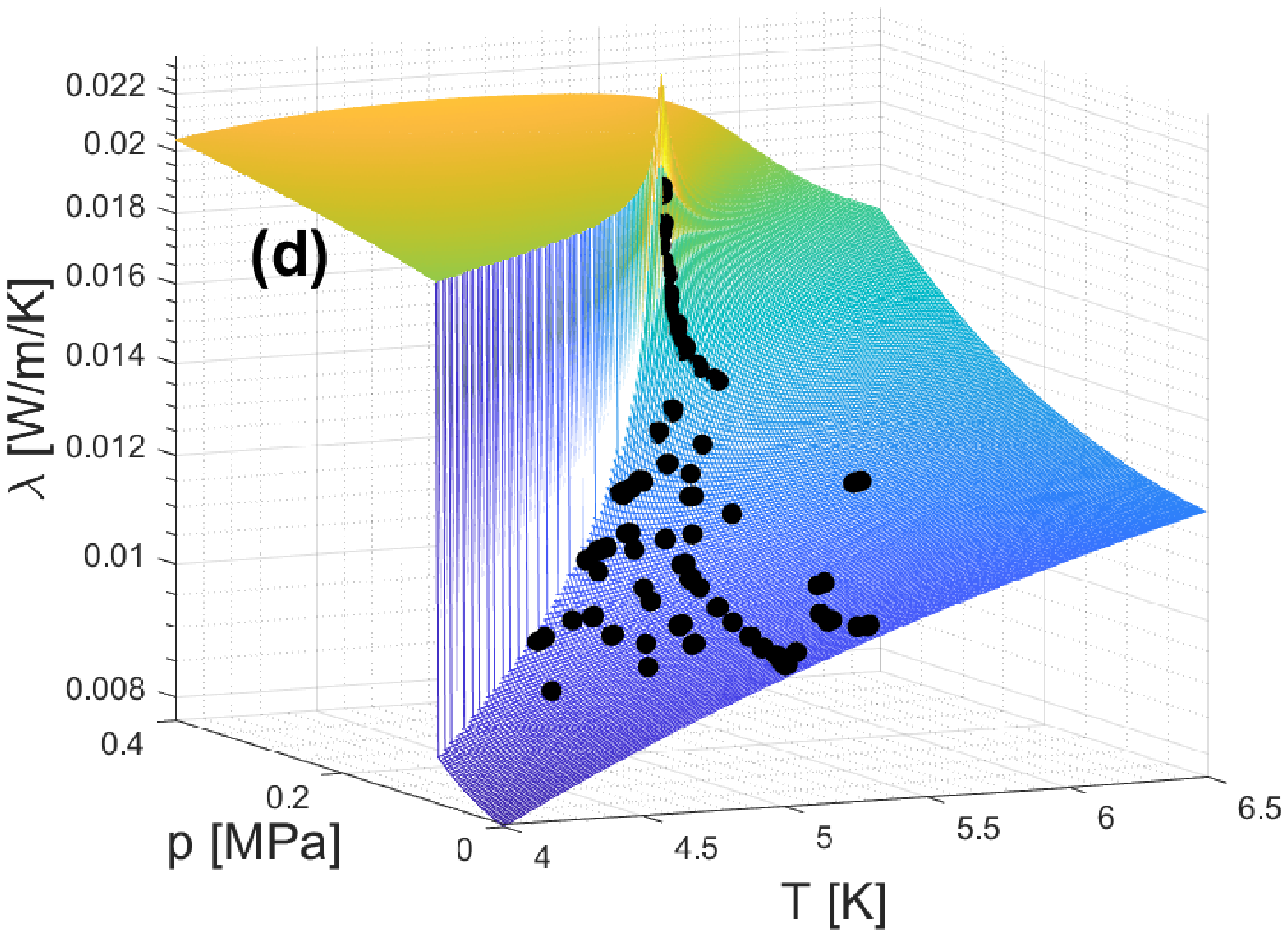}
\caption{Basic state quantities (a,b) and molecular transport properties (c,d) which characterize the complex material properties of cryogenic helium $^4$He near its critical point (CP) of $p_{\rm CP} = 227$ kPa and $T_{\rm CP} = 5.2$ K as a function of pressure $p$ and temperature $T$.  Panel (a) shows the density $\rho$ which thus corresponds with the equation of state $\rho(T,p)$. Panel (b) plots the specific heat $C_p$, panel (c) the dynamical viscosity $\mu$, and panel (d) the thermal conductivity $\lambda$. The black bullets indicate the mean values $p_m$ and $T_m$ for all experiments shown in Fig~\ref{fig:PTexperiments}. The data are obtained from the HEPAK code~\cite{Arp1998}.}
\label{fig:Surfaces}
\end{figure*}

Figure \ref{fig:Surfaces} shows the accurate values of the mass density $\rho$, the specific heat at constant pressure $C_p$, the dynamic viscosity $\mu$, and the thermal conductivity $\lambda$ with respect to the $p$--$T$ plane in a region of gaseous helium phase including the regions near to the vapor liquid saturation curve and the critical point. These surfaces are obtained from the HEPAK code~\cite{Arp1998}, which is based on high-order interpolation of all available and reliable measurements. The plots give us a first guidance in the selection of the appropriate the functional form of the state and transport properties in the equations beyond the OB approximation. All displayed quantities, $\rho$, $C_p$, $\lambda$, and $\mu$ develop a discontinuity at the phase boundary in the pressure-temperature plane which corresponds to a first-order gas--liquid phase transition. Furthermore, $C_p$ and $\lambda$ develop a divergence at the critical point (CP), which is given by $p_{\rm CP} = 227$ kPa and $T_{\rm CP} = 5.2$ K. The precursors of this divergence are visible here. Panel (a) displays in fact the equation of state $\rho(T,p)$. In the vicinity of this point, most of the RBC experiments in cryogenic helium have been performed as indicated by the black dots in all panels of the figure. They indicate mean pressure and temperature (and a view from the top would reproduce the points of Fig. \ref{fig:PTexperiments}). Notice that the density $\rho$ and the specific heat $C_p$ vary by several orders of magnitude over the domain displayed, while the dynamic viscosity $\mu$ and the thermal conductivity $\lambda$ vary by a factor of smaller than 4 only. 

Finally, in order to assess possible contribution of compressibility to the NOB effects, we evaluate the Mach number $M = U_{\rm{f}} / c = \sqrt{g \alpha\Delta T H} / c$, with isobaric thermal expansivity $\alpha$ and the speed of sound $c$ obtained from HEPAK (not shown). The results show that for all experiments considered, the Mach numbers are in the range $M \lesssim  10^{-2}$ and guarantee that possible breaking of OB conditions due to compressibility can be neglected. Thus the NOB effects in RBC experiments with cryogenic helium stem solely from temperature dependencies of the fluid properties.

\section {Experiments in cryogenic $^4$He}\label{sec:Apparatus}
Cryogenic helium $^4$He has been used to reach extreme turbulence intensity in ``tabletop'' RBC experiments on one hand thanks to the peculiar material properties near the CP allow to reach high Rayleigh numbers $Ra$, on the other hand, due to technical advantages, as heat leaks and many other parasitic effects are naturally highly suppressed in cryogenic conditions. The first advantage goes side-by-side with caveats of inevitably varying the Prandtl number $Pr=\nu/\kappa$ \cite{Niemela2000,Urban2011} as well as NOB effects stemming from (\ref{eq:TildeD}) and the dependencies \eqref{Mat1}--\eqref{Mat5}. Here we re-analyze RBC data obtained at the Brno cryogenic turbulence facility \cite{Urban2010}. 

The Brno experiment comprises a cryostat with a helium cryogenic experimental cell with the height $H = 0.3$ m and diameter $d = 0.3$ m (aspect ratio $\Gamma = d/H = 1$) with particular effort to minimize the influence of the cell structure and materials on the observed convection. The cell has been designed to withstand pressures of $3.5$ bars to cover a range of Rayleigh numbers $10^7\le Ra\le 10^{15}$. Here, we list the main features of the experimental cell only, including all crucial recent upgrades. In an ideal RBC experiment, the top and bottom plates should maintain non-fluctuating constant temperatures $T_{\rm top}$ and $T_{\rm bot}$ and the cell sidewalls should be adiabatic. The top and bottom plates of our cell are made of 28 mm thick annealed OFHC copper of very high thermal conductivity of at least $\lambda_{\rm Cu}=2$ kW m$^{-1}$ K$^{-1}$ at 5 K. Parasitic heat fluxes from the sidewalls into the working fluid are minimized by using very thin stainless steel sidewalls with thickness $\delta = 0.5$ mm and a thermal conductivity $\lambda_w$. A special design of the cell corners is used, see Fig. 4 of~\cite{Urban2010}. One way to estimate the influence of the sidewall on the heat transport is via the wall parameter $W = 4\delta \lambda_w/(\lambda_{\rm He} D)$; for our cell $0.22 > W >0.15$ depending on actual value of the thermal conductivity for each data point. By correction, we mean a subtraction of the heat conducted by sidewalls from the heat that passes through the working fluid by convection~\cite{Roche2001}. We also paid attention to employ good thermal shielding and minimize other external parasitic heat flows into the cell which could substantially influence the convection dynamics.

A temperature correction which needs to be addressed in cryogenic experiments is that due to adiabatic gradient, $g/C_p$. It is given by $\Delta T_{\rm{ad}} = g \alpha H T_m / C_p = \tilde{D} T_m$, see also Eq.~(\ref{eq:TildeD}), which has to be subtracted from the measured temperature difference $\Delta T$ before evaluating $Ra$ and comparing to the results of DNS based on Eqs.~\eqref{eq:Cont}--\eqref{eq:Heat}, otherwise the experimental data points would be very much off the expected $Nu(Ra)$ dependencies. In typical large RBC cryogenic helium experiments, including the experiments in Brno, $\Delta T_{\rm{ad}}$ is of order 1 mK. In the largest RBC cell at Oregon, it was about 3 mK~\cite{Niemela2000}. In contrast, in turbulent RBC in air at room temperature the adiabatic gradient correction is not important, as $g/C_p=0.01$ K/m; thus the scale height across which $T$ drops by an order of magnitude is 1 km.

Cryogenic helium does not absorb thermal radiation, which leaves the radiation corrections to the Nusselt number negligibly small. We point to a straightforward evaluation in ref.~\cite{Urban2020rad} and to the discussion of an extreme case in~\cite{Lepot18}. 

The top and bottom plates of the cell are equipped with four germanium (Ge) thermometers, calibrated at Physikalisch-Technische Bundesanstalt Berlin (Germany) up to the best currently available precision of $\pm 2$ mK absolute accuracy over the entire temperature range of interest. These Ge thermometers are embedded in the middle and on the sides of both plates. We see no horizontal temperature gradients within Ge sensors accuracy of 2 mK in both copper plates. Following the last upgrade, both copper plates contain a pair of fast response Lakeshore DT-670 silicon (Si) diode thermometers, allowing to resolve the plate temperature fluctuations, which on one hand enable much better control of the temperature boundary conditions~\cite{Urban2021} compared to previous experiments via Proportional-Integrative-Derivative feedback loops and, on the other hand, can be used to correlate dynamics of the turbulent large-scale circulation in the bulk with temperature fluctuations affecting the boundary layers. Additional sensors are placed inside the cell, calibrated in situ by us against the primary four Ge thermometers, and can be adjusted to measure directly the absolute turbulent core temperature $T_c$ or the fluctuations $\delta T_c$. 

\section {Nonlinear regression of non-Boussinesq parameters} \label{sec:ML}
\subsection{Experimental data base and regression procedure}

We are interested in the dependence of the NOB parameter on state variables and material properties, i.e., most generally this results to a function $\xi(p_m, T_m, \Delta T, \alpha(T,p), \lambda(T,p), \mu(T,p), C_p(T,p), \rho(T,p))$. The perturbative expansion of Sec. II has reduced this high-dimensional function to one of 13 input parameters. These are the mean pressure $p_m$, the mean temperature $T_m$, the temperature difference between bottom and top boundaries $\Delta T$, and the expansion parameters $\varepsilon_i$ for $i=1,...,10$. The latter ones determine the first and second temperature derivatives of the material properties and state variables as summarized in Table \ref{Tab_eps}. The nonlinear regression proceeds in three levels of increasing complexity. We aim at reconstructing the following functions, 
\begin{align}
\xi_1 &:= \xi_{\rm NOB}(p_m, T_m, \Delta T)\,,
\label{xi1}\\
\xi_2 &:= \xi_{\rm NOB}(p_m, T_m, \Delta T, \varepsilon_{2k+1}) \quad \mbox{for} \quad k=0, ..., 4\,,
\label{xi2}\\
\xi_3 &:= \xi_{\rm NOB}(p_m, T_m, \Delta T, \varepsilon_{k}) \quad \mbox{for} \quad k=1, ..., 10\,.
\label{xi3}
\end{align}
In the first approach $\xi_1$, no expansion parameter $\varepsilon_i$ is used for the training of the neural network. It is explored how the NOB parameter depends on mean pressure and temperature as well as imposed temperature difference only. The result can serve as a baseline to forecast NOB effects on the temperature profile asymmetry $\xi_{\rm NOB}(p_m, T_m, \Delta T)$ in future experiments. In the more detailed successive steps, we are interested in a finer resolution of the effects of individual fluid properties on the temperature asymmetry. In the second (third) approach $\xi_2$ ($\xi_3$), sets of all linear-order (linear and quadratic-order) expansion coefficients $\varepsilon_{2k+1}$ ($\varepsilon_{2k}$ ) are taken together as input data for a slightly deeper neural network architecture since the feature extraction proceeds in a higher-dimensional feature space. In this case, before calculating the non-Boussinesq parameter, it was necessary to perform preliminary calculations to determine $\varepsilon_k$ at each point of the $p$--$T$-space within the given $\Delta T$ by means of HEPAK. Values of the NOB parameters (\ref{xi1})-(\ref{xi3}) are given in units of K throughout the paper.

\begin{figure}[h]
\centering
\includegraphics[width=0.95\linewidth]{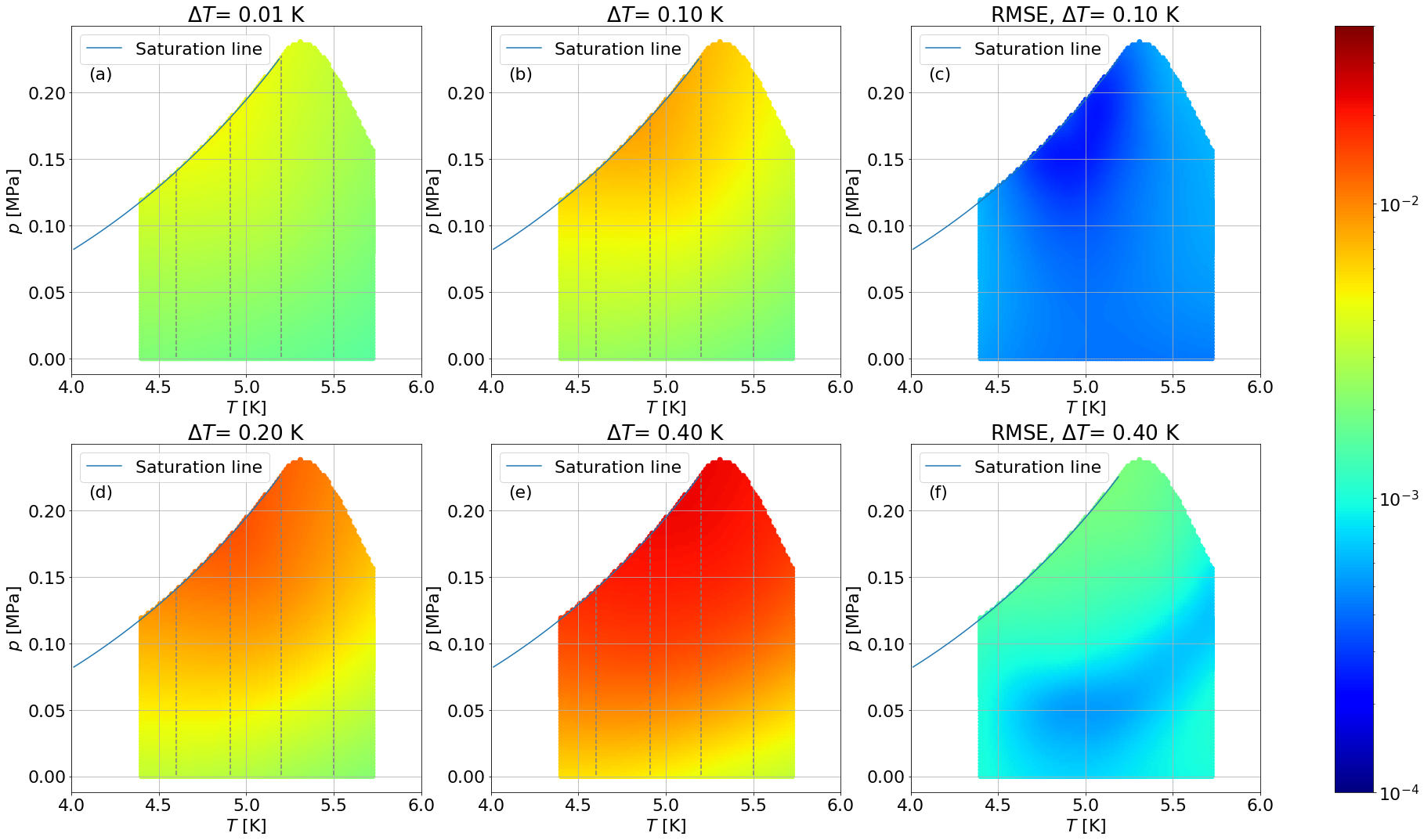}
\caption{Contour plots of the reconstruction of the non-Boussinesq parameter $\xi_1(p,T)$ for different $\Delta T$ is shown in panels (a), (b), (d), and (e). Panels (c) and (f) display the corresponding root mean square error (RMSE) for two of the four cases. The RMSE is given by Eq.~\eqref{rmse1}. Unit for the color code is Kelvin.}
\label{Tm-Tc_0}
\end{figure}

\subsection{First reconstruction method without expansion parameters}\label{rec1}

In the first reconstruction method, there are three input parameters, the mean pressure $p_m$ and the mean temperature $T_m$ which define the operating point of a particular laboratory measurements. The third input parameter is the applied temperature difference between the bottom and top plates. The only output parameter is the difference between mean temperature $T_m$ and temperature $T_c$ in the center of the layer, $\xi_1$, as already defined in the last subsection, see Eq.~\eqref{xi1}. For details on the network architecture, see table \ref{Tab1_app} in Appendix A. 

Figure \ref{Tm-Tc_0}, shows the reconstruction of $\xi_1(T,P)$ for 4 different outer temperature differences $\Delta T$ which are indicated in the title of panels (a), (b), (d), and (e). In panels (c) and (f), we display in addition the root mean square error (RMSE) contour plots for two cases which arises when taking the $i=1, ..., N_{\rm rec}=100$ individual reconstructions. It is defined as 
\begin{equation}
{\rm RMSE}(p,T)=\sqrt{\frac{1}{N_{\rm rec}}\sum_{i=1}^{N_{\rm rec}} |\xi_k(i,T,p)-\overline{\xi}_k(T,p)|^2} \quad \mbox{for}\quad k=1, 2, 3\,, 
\label{rmse1}
\end{equation}
where $\overline{\xi}_k(T,p)$ the mean reconstructed surface. In panels (a), (b), (d), and (e), it can be seen that the maximum value of the NOB parameter is always reached in the vicinity of the phase boundary and at the critical point. As the temperature difference $\Delta T$ increases, the NOB parameter also increases. In panels (c) and (f), it is seen that the RMSE also increases for bigger $\Delta T$. One reason is that there are less corresponding experimental data. The largest errors occur near the critical point at the end of the phase boundary.

\begin{figure}[h]
\centering
\includegraphics[width=0.5\linewidth]{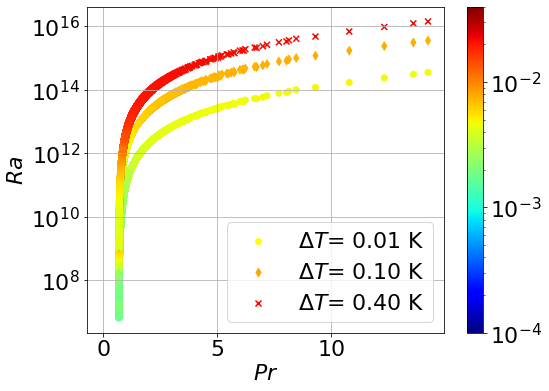}
\caption{Plot of $\xi_1$ in the $Ra$--$Pr$ parameter space for three different outer temperature differences $\Delta T$. The data points are color-coded by $\xi_1$ as given by the legend to the right. Unit for the color code is Kelvin.}
\label{Ra-Pr_0order}
\end{figure}

The non-Oberbeck-Boussinesq parameter $\xi_1$ is also shown in $Ra$--$Pr$ parameter space in Fig. \ref{Ra-Pr_0order} for three outer temperature differences $\Delta T$, cf. Fig.~\ref{fig:PTexperiments}(a) for an analogous display of experimental values. All branches for each $\Delta T$ start at $Pr=0.7$ for the smallest Rayleigh numbers. When the Rayleigh number increases by 8 orders of magnitude, the Prandtl number {increases monotonically up to $Pr \simeq 15$}. The data points are color-coded by $\xi_1$. The NOB parameter also grows along the curves up to its highest values at $Pr\simeq 2$. With a further increase of the Prandtl number, the NOB parameter $\xi_1$ remains however nearly unchanged.

\subsection{Second reconstruction method including linear-order $\varepsilon$-parameters}\label{rec2}

For the next level, we reconstruct the parameter field $\xi_2$ with the linear order expansion parameters $\varepsilon_i$. Input parameters for the deep neural network are now mean pressure $p_m$, mean temperature $T_m$, outer temperature difference $\Delta T$ (as in the last case) together with all $\varepsilon_{2k+1}$ for $k=0, ..., 4$. The output is now the NOB parameter $\xi_2$ of Eq.~\eqref{xi1}. The neural network is detailed in table \ref{Tab2_app} of Appendix A.
  
The results are shown in Fig. \ref{fig:Tm_Tc_1order}, in analogy with Fig. \ref{Tm-Tc_0}, for four different temperature differences. Also, we add again two root mean square error plots in panels (c, f) for $\Delta T = 0.1$ and $0.4$, respectively. It can be seen that the qualitative behavior is similar to the results of the first reconstruction method. The difference between the reconstructions $\xi_1$ and $\xi_2$ is highlighted in Fig.~\ref{2nd-1st}, showing the deviation $|\xi_1-\xi_2|$. Maximum deviations are up to $\xi\simeq 0.005$ K, observed for $\Delta T = 0.2$ K near the critical point and for $\Delta T = 0.4$ K in the lower-$T$ part near the saturation curve (maximum relative value of $|\xi_1 - \xi_2| / \xi_1 \sim 40\%$). The graphs show that the influence of linear coefficients near the critical point grows as the temperature difference increases.

The $Ra$--$Pr$ plots related to the $|\xi_1 - \xi_2|$ differences are shown in Fig. \ref{Ra_Pr_1_order} and display the largest deviation near the critical point and at the saturation line when $Pr \gtrsim 2$ and $Ra$ varying from $10^{12}$ to $10^{14}$.
\begin{figure}[h]
\centering
\includegraphics[width=0.95\linewidth]{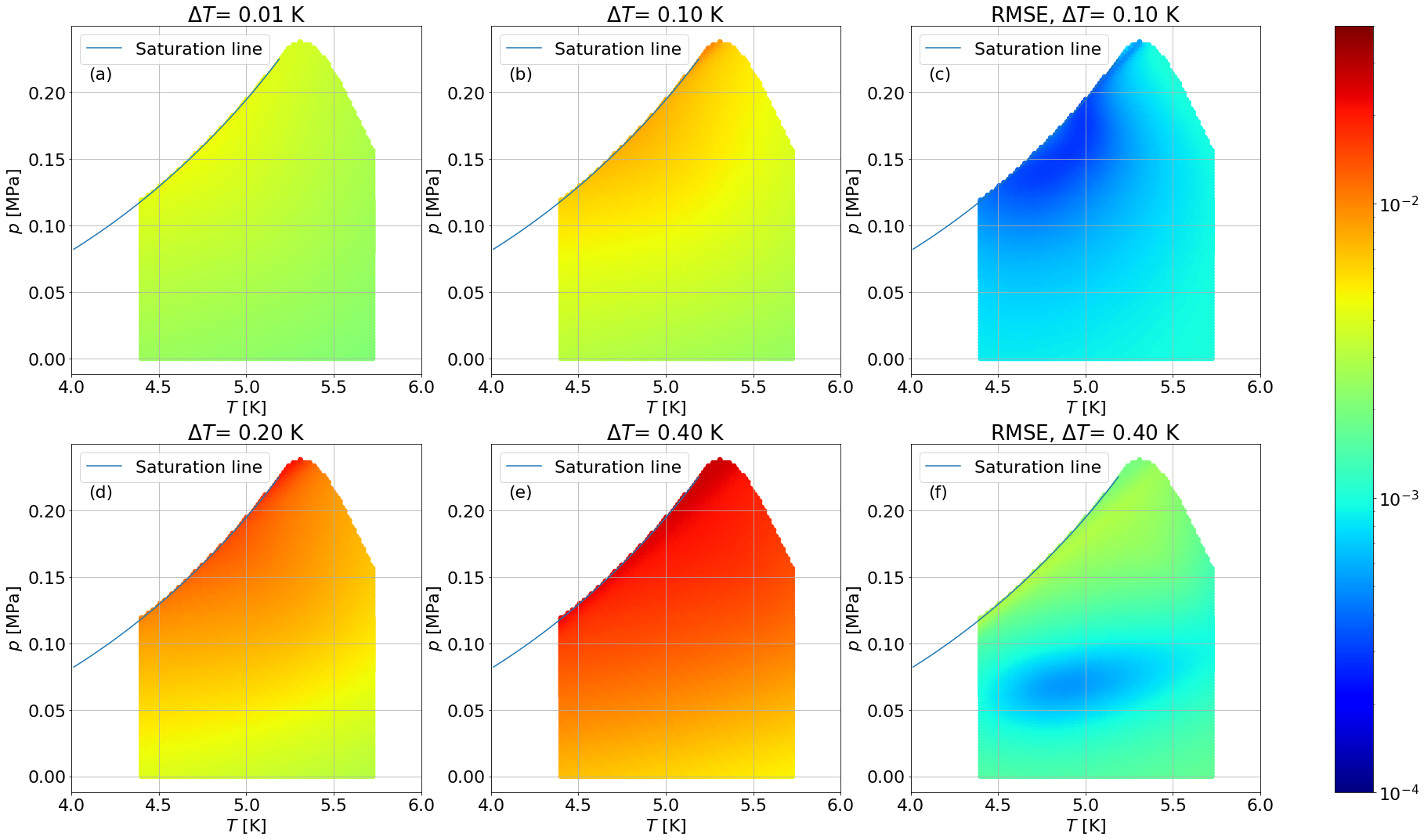}
\caption{Contour plots of the reconstruction of the non-Boussinesq parameter $\xi_2(p,T)$ for different $\Delta T$ is shown in panels (a), (b), (d), and (e). Panels (c) and (f) display the corresponding root mean square error (RMSE) for two of the four cases. The RMSE is given by Eq.~\eqref{rmse1}. The unit for the color code is Kelvin.}
\label{fig:Tm_Tc_1order}
\end{figure}
\begin{figure}[h]
\centering
\includegraphics[width=0.75\linewidth]{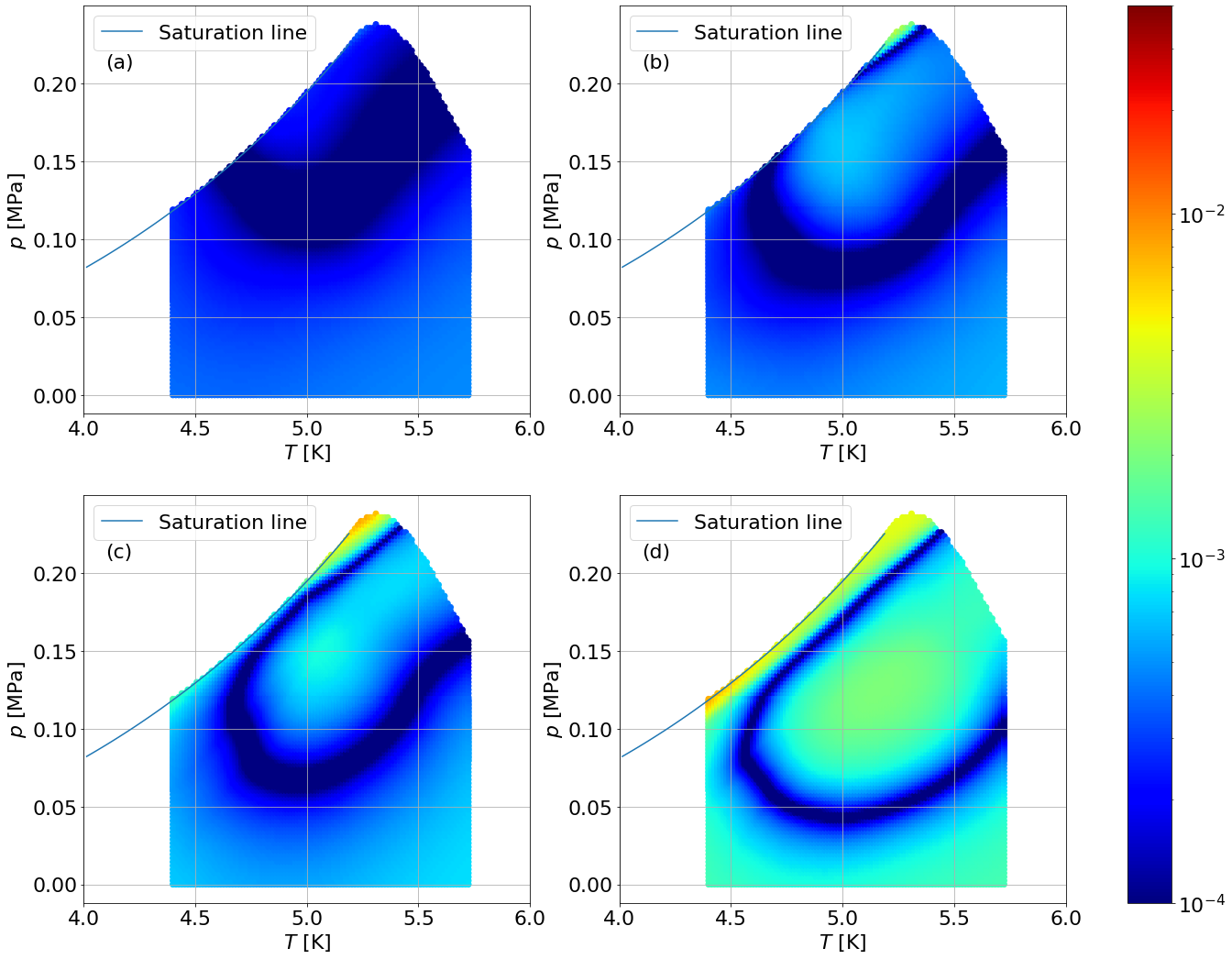}
\caption{Contour plots of the absolute difference $|\xi_1-\xi_2|$ between the 1st and 2nd reconstruction method in the $p$--$T$ parameter plane for four different $\Delta T$ values which correspond to those in panels (a), (b), (d), and (e) of Figs. \ref{Tm-Tc_0} and \ref{fig:Tm_Tc_1order}. The unit for the color code is Kelvin.}
\label{2nd-1st} 
\end{figure}
\begin{figure}
\includegraphics[width=0.46\textwidth]{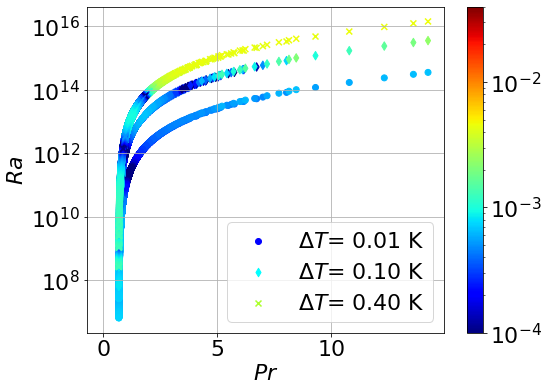} 
\caption{The absolute difference $|\xi_1 - \xi_2|$ between the first and the second reconstruction method in the $Ra$--$Pr$ parameter space for three different outer temperature differences $\Delta T$ (cf. Fig.~\ref{2nd-1st}). The unit for the color code is Kelvin.}
\label{Ra_Pr_1_order}
\end{figure}

\subsection{Third reconstruction method including quadratic-order $\varepsilon$-parameters}\label{rec3}

\begin{figure}[h]
\centering
    \includegraphics[width=0.495\linewidth]{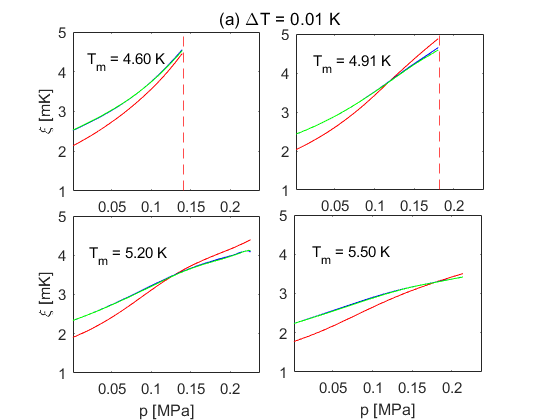}
    \includegraphics[width=0.495\linewidth]{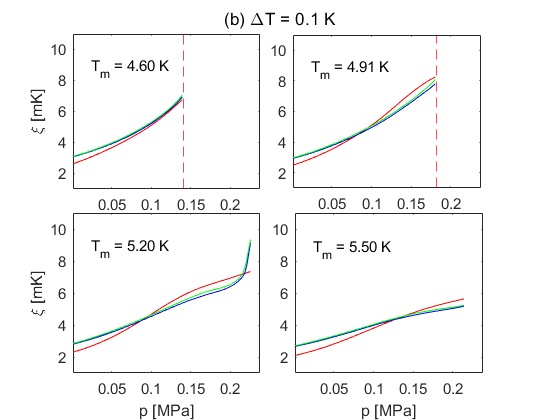}
    \includegraphics[width=0.495\linewidth]{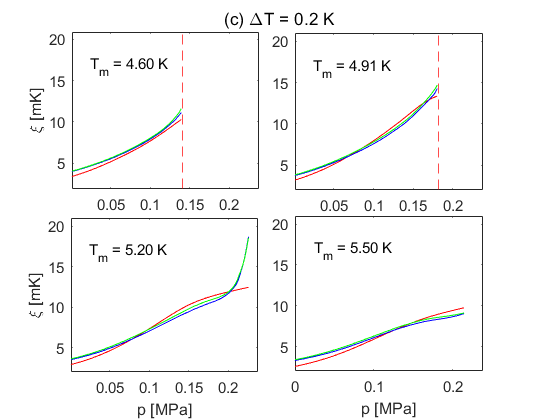}
    \includegraphics[width=0.495\linewidth]{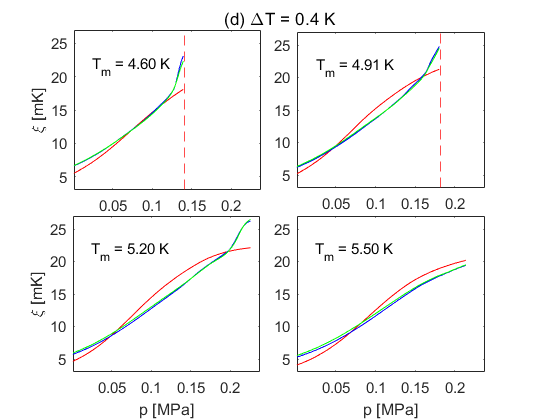}
\caption{Comparison of individual reconstructions of the NOB parameter $\xi_1$ (red), $\xi_2$ (blue) and $\xi_3$ (green) plotted as functions of pressure $p$ at four values of $T_m=4.6$, 4.91, 5.2, and 5.5 K (see insets in panels) and four values of $\Delta T = 0.01$, 0.1, 0.2, and 0.4 K (cf. Figs.~\ref{Tm-Tc_0} and \ref{fig:Tm_Tc_1order}). The vertical red dashed lines in panels with $T_m = 4.6$~K and $4.91$~K denote the $p$ values at the SVC.}
\label{Tm_Tc_1order1D}
\end{figure}

The last reconstruction method includes 13 input parameters to obtain $\xi_3$. This comprises linear and quadratic expansions with respect to the temperature encoded by $\varepsilon_i$ for $i=1, \dots, 10$. Our analysis shows that the absolute difference $|\xi_2-\xi_3|$ remains very small for all the $p_m$, $T_m$, $\Delta T$ values considered. This can be seen in Fig.~\ref{Tm_Tc_1order1D}, which corresponds to four 'sections' through the $p-T$-plane at $T_m = 4.6, 4.91, 5.2$ and $5.5$ K showing the $p$-dependencies of $\xi_1$, $\xi_2$ and $\xi_3$. Finally, we refer to Appendix B, where we have summarized further error analysis for all three reconstruction methods. This analysis illustrates how strongly the 100 individual reconstructions of $\xi_k$ vary when selecting different subsets as training and test data, see Fig. \ref{calc1}.   

\begin{figure}
    \centering
\includegraphics[width=0.45\textwidth]{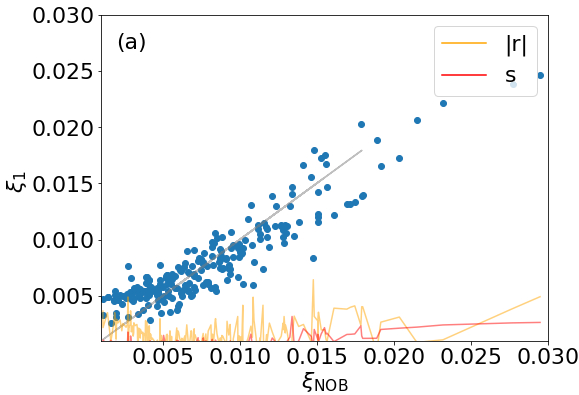} 
\includegraphics[width=0.45\textwidth]{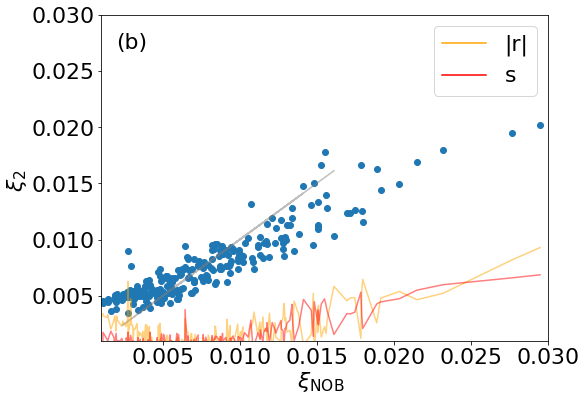} 
\includegraphics[width=0.45\textwidth]{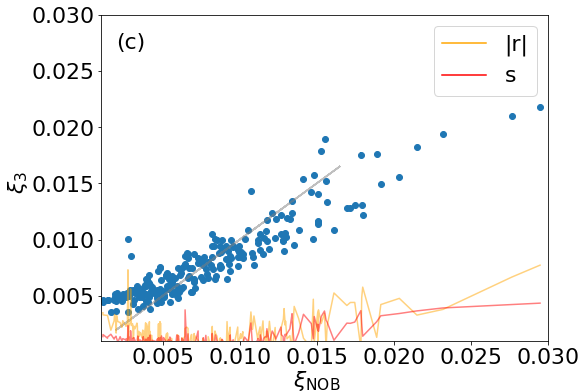} 
    \caption{Correlation of the reconstructed values of the NOB parameters $\xi_1$ (a), $\xi_2$ (b) and $\xi_3$ (c) of Eqs.~(\ref{xi1})-(\ref{xi3}) with the experimental values of $\xi_{\rm NOB}$ of Eq.~(\ref{ksiNOB}). The gray line is the ideal fit for visual guidance. The red and yellow curves at the bottom of each panel represent the standard deviation $s$ and the sum of residuals $|r|$ for the ensembles of $N_{\rm rec} = 100$ neural network runs to obtain each point, respectively; see Appendix B for details.}
    \label{fig:my_label} 
\end{figure}

\subsection{Discussion of the reconstruction methods}
The results in Secs.~\ref{rec1}--\ref{rec3} display a general trend for the magnitude of NOB effects in RBC in cryogenic helium gas quantified by the response parameters $\xi_i$, $i=1,2,3$. All of them grow significantly as we increase pressure $p_m$ and decrease temperature $T_m$ towards the phase boundary at SVC and towards the CP. This is seen throughout Figs.~\ref{Tm-Tc_0} and \ref{fig:Tm_Tc_1order} for different $\Delta T$ values, chosen here to cover the ranges of $\Delta T$ taken in high-$Ra$ turbulent RBC experiments. Figure~\ref{Tm_Tc_1order1D} shows in more detail the differences between individual reconstructions obtained by different neural network architectures. We can observe that while the differences between $\xi_2$ and $\xi_3$ are practically negligible, the reconstruction calculating $\xi_1$, involving only the basic experimental parameters, namely $p_m$,$T_m$ and $\Delta T$, significantly differ from $\xi_2$ and $\xi_3$, which in addition take into account the linear and quadratic terms in the expansion of the material properties of helium as a function of temperature, expressed by the dimensionless NOB control parameters $\varepsilon_i$, $i=1,...,10$, given in Tab.~\ref{Tab_eps}. In particular, $\xi_2$ and $\xi_3$ show much more pronounced growth at the phase boundary near the SVC and at the CP as a function of pressure than $\xi_1$. Further, the $\xi_2$ and $\xi_3$ curves are on average markedly more convex, while $\xi_1$ often grows in a more concave fashion towards the SVC and/or CP, see e.g. the most pronounced case for $\Delta T = 0.2$~K and $T_m = 5.2$~K (very close to the critical temperature $T_{\rm cri} = 5.195$~K). Note that the occurrence of crossings of the concave regions of $\xi_1$ with convex parts of $\xi_2$ apparent in Fig.~\ref{Tm_Tc_1order1D} explains appearance of tongue-shaped features (contours of $|\xi_1-\xi_2$ =0) seen in Fig.~\ref{2nd-1st}.

In addition to the three reconstructions (\ref{xi1})-(\ref{xi3}) detailed in Secs.~\ref{rec1}--\ref{rec3}, we performed also several four-parameter reconstructions taking individual $\varepsilon_i$, $i=1,...,10$ one-by-one in addition to $p_m$, $T_m$ and $\Delta T$ within the appropriate neural network. Each of the results differed from the $\xi_1$ surfaces shown in Fig.~\ref{Tm-Tc_0} by less then the experimental accuracy of $~3$mK. Thus only the combination of the linear expansion parameters, resulting in $\xi_2$ of Eq.~(\ref{xi2}), can be considered significant.

In Fig.~\ref{fig:my_label}, we finally plot the $\xi_{\mathrm{NOB}}$ results obtained by the machine learning ($y$-axis) in comparison to the experimental data ($x$-axis) for $\xi_1$ (a), $\xi_2$ (b), and $\xi_3$ (c). Each point thus corresponds to an experimental value and a value obtained from the neural network. In addition, a line $y=x$ for an ideal fit to the data is shown. The presented spread of points shows a fairly good reconstruction of the non-Boussinesq parameters by the ML algorithm.

\section {Conclusion and outlook}\label{sec:SumCon}
In this work, we investigated the non-Oberbeck-Boussinesq behavior in high-Rayleigh-number laboratory experiments of turbulent Rayleigh-B\'{e}nard convection in cryogenic helium $^4$He. The NOB effects in this experimental setup are shown to be caused by the temperature dependence of the material properties at the molecular level, while the compressibility effects can be neglected with a Mach number of $M \lesssim 10^{-2}$. The temperature $T_c$ measured at the center of the RBC cell is found to deviate from the arithmetic mean of the temperatures of the copper plates at the top and bottom, $T_m$. This is an indicator of an asymmetry of the statistical properties between the top and bottom in the cell, which unambiguously signifies breaking of the OB approximation. The deviations have been determined in a series of experiments which provide a sparse data set to reconstruct the function $\xi_{\rm NOB}(p,T)=T_m-T_c$ by a nonlinear regression. The experiments are characterized by the operating point in the pressure--temperature plane, $(p_m, T_m)$, and the outer temperature difference, $\Delta T$. We thus provide a smooth approximation (at different levels of accuracy) for the strength of the NOB effects with respect to two state variables. 

In detail, we performed reconstructions of the NOB parameter by deep neural networks in three different ways. The first approach is based on the operating point $p_m, T_m$ and $\Delta T$. The second and third approaches incorporate the expansion coefficients up to the linear and quadratic orders of the Taylor expansion with respect to $\delta T=T-T_m$ of the material properties and mass density, respectively. The comparison of the different methods can be summarized as follows: (i) The inclusion of the linear-order temperature expansion alters the reconstruction results by up to 40\%. (ii) The inclusion of the second expansion order does not alter the magnitude of the NOB effects significantly. Our study provides a first systematic reconstruction of the NOB effects in experiments with cryogenic helium, and renders a set of maps for expected NOB effects in a wide area of the pressure--temperature plane for different values of $\Delta T$ in future experiments. A systematic analysis of the impact of the expansions on the heat and momentum transfer could be a next step which would require numerical investigations of the OB configuration for these parameters.

\acknowledgements
The joint project is supported by grant no. 21-06012J of the Czech Science Foundation (GA\v{C}R) for M.M. and by grant no. SCHU 1410/31-1 of the Deutsche Forschungsgemeinschaft (DFG) for G.Z. The work of G.Z. is co-funded by the European Union (ERC, MesoComp, 101052786). Views and opinions expressed are however those of the authors only and do not necessarily reflect those of the European Union or the European Research Council. Neither the European Union nor the granting authority can be held responsible for them. Finally, we thank Ladislav Skrbek, Olga Shishkina, and Valentina Valori for discussions.

\appendix
\section{Details of the deep neural networks for nonlinear regression}
\label{Appendix_A}

In the following, we will provide some technical details of the machine learning methods that were used for the nonlinear regression analysis, see e.g. \cite{Goodfellow2016}. In order to calculate all results, we had to use network models with different input and output parameters. They are provided in the following three tables together with some information. In the 1st model we trained a neural network without any information about the perturbative expansion as explained in the main text. The model consists of a total of 6 layers. These are (1) linear layers which apply a linear transformation of the incoming data followed by (2) a rectified linear unit (ReLU) function which performs an element-wise nonlinear activation. ReLU  and the parametric ReLU (PReLU) are given by 
\begin{equation}
{\rm  ReLU}(x)=\max_x (0, x)\quad\mbox{and}\quad {\rm PReLU}(x)=\max (0, x)+a\min(0,x)\,,
\end{equation}
for input data $x$ with parameter $a>0$. Furthermore, (3) the BatchNorm1d function of the PyTorch library is applied as a batch normalization to fix expectation value E and variance Var of the input in order to accelerate the training. This function is given by 
\begin{equation}
{\rm  BatchNorm1d}(x)=\frac{x-{\rm E}[x]}{\sqrt{{\rm Var}[x]+\epsilon}}\,,
\end{equation}
for input data $x$ in the form of a mini-batch. Finally, (4) the sigmoid applies another element-wise nonlinear activation which is given by
\begin{equation}
\sigma(x)=\frac{1}{1+\exp(-x)}\,.
\end{equation}
This first neural network in table \ref{Tab1_app} obtains $\xi_1$ as the output resulting from three quantities at input, see Eq.~\eqref{xi1}.
\begin{table}[h!]
\begin{tabular}{lcc}
\hline\hline
Layer  & Output shape & $\quad$Number of parameters$\quad$ \\ \hline
Linear & [16, 400] & 1600 \\ 
ReLU & [16, 400] & 1 \\ 
BatchNorm-1d & [16, 400] & 800 \\
Linear & [16, 1] & 401 \\ 
BatchNorm-1d & [16, 1] & 2 \\ 
Sigmoid & [16, 1] & 0 \\ \hline
Total & ~ & 2804 \\ \hline\hline
\end{tabular}
\caption{Details of the deep neural network which is used for the first approach. The output shape consists of the batch size and the number of weights.}
\label{Tab1_app}
\end{table}
For our nonlinear regression task, there are however only 236 data points available to train the neural network. This is a small number for a full training of a deep neural network in the considered pressure-temperature interval. It can lead to the fact that in the randomly chosen input data for the training-testing procedure, no data point near the critical line/point is chosen at all. This in turn can result in an incorrect approximation of the non-Boussinesq parameter $\xi_{\rm NOB}$. In order to remove this effect of an insufficient amount of experimental data, training and testing were performed in a cycle of 100 runs at each of the three levels. Each run takes a  randomly chosen subsample of the RBC data and uses the rest for testing. The deviation of each calculation from the mean is shown in appendix \ref{Appendix_B}. In this way, we obtained more robust regression results for a lacking number of RBC experiments.

The second reconstruction method includes the parameters $\varepsilon$ of the linear order expansion. These parameters, however, cannot be obtained close to the phase boundary from HEPAK and were thus reconstructed first by a neural network. The corresponding architecture is specified in the right fraction of table \ref{Tab2_app}. Input is again $p_m$, $T_m$, and $\Delta T$. With these input parameters, we obtain 5 outputs, namely $\varepsilon_1$, $\varepsilon_3$, $\varepsilon_5$, $\varepsilon_7$, and $\varepsilon_9$. Subsequently, we add them as further input parameters, i.e., we have a total of 8 inputs to calculate the contour plot of $\xi_2$ in the $p$--$T$ plane. The recursion is now more complicated in comparison to the first approach because all property gradients in the experimental data fluctuate strongly, particularly close to the phase boundary. Also, the model is more complex since it includes more parameters. The architecture of the neural network is basically the same for both substeps. 

\begin{table}[h!]
\begin{tabular}{lccccc}
\hline\hline
~&\multicolumn{2}{c}{Reconstruction of $\xi_2$} & $\quad$ & \multicolumn{2}{c}{Evaluation of $\varepsilon_{2i+1}$}\\
Layer & Output shape & $\quad$Number of parameters$\quad$ & $\quad$ & Output shape & $\quad$Number of parameters$\quad$ \\ \hline
Linear & [16, 400] & 3600 & $\quad$ & [16, 500] & 2000 \\ 
PReLU & [16, 400] & 1 & $\quad$ &[16, 500] & 1 \\ 
BatchNorm-1d & [16, 400] & 800 & $\quad$ & [16, 500] & 1000 \\ 
Linear & [16, 1] & 401 & $\quad$ & [16, 5] & 2505 \\ 
BatchNorm-1d & [16, 1] & 2 & $\quad$ & [16, 5] & 10 \\ 
Sigmoid & [16, 1] & 0 & $\quad$ & [16, 5] & 0 \\ \hline
Total  & ~ & 4804 & $\quad$ & ~ & 5516 \\ 
\hline\hline
\end{tabular}
\caption{Details of the deep neural network which is used for the second approach. The output shape consists of the batch size and the number of weights for each network layer.}
\label{Tab2_app}
\end{table}
The third reconstruction method is similar to the second. Now, we use linear and quadratic orders, reconstruct the 10 expansion coefficients first, and obtain $\xi_3$ from a network with 13 inputs. The architecture is specified in table \ref{Tab3_app}. It is basically similar to the second reconstruction approach. 
\begin{table}[h!]
\begin{tabular}{lccccc}
\hline\hline
~&\multicolumn{2}{c}{Reconstruction of $\xi_3$} & $\quad$ & \multicolumn{2}{c}{Evaluation of $\varepsilon_i$}\\
Layer & Output shape & $\quad$Number of parameters$\quad$ & $\quad$ & Output shape & $\quad$Number of parameters$\quad$ \\ \hline
Linear & [16, 400] & 5600 & $\quad$ & [16, 500] & 2000 \\ 
PReLU & [16, 400] & 1 & $\quad$ & [16, 500] & 1 \\ 
BatchNorm-1d & [16, 400] & 800 & $\quad$ & [16, 500] & 1000 \\ 
Linear & [16, 1] & 401 & $\quad$ & [16, 10] & 5010 \\
BatchNorm-1d & [16, 1] & 2 & $\quad$ & [16, 10] & 20 \\ 
Sigmoid & [16, 1] & 0 & $\quad$ & [16, 10] & 0 \\ \hline
Total & ~ & 6804 & $\quad$ & ~ & 8031 \\ 
\hline\hline
\end{tabular}
\caption{Details of the deep neural network which is used for the third approach. The output shape consists of the batch size and the number of weights.}
\label{Tab3_app}
\end{table}

\section{Error analysis of the nonlinear regression}
\label{Appendix_B}

\begin{figure}[h!]
\includegraphics[width=0.45\textwidth]{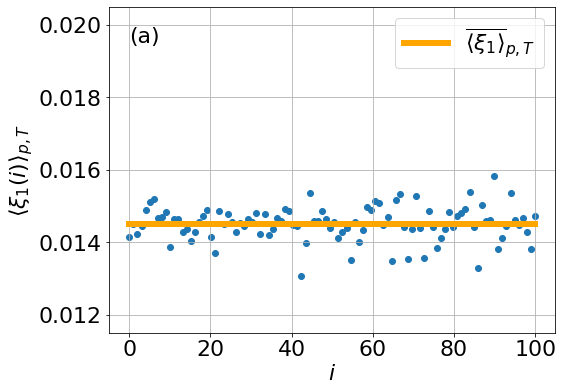} 
\includegraphics[width=0.45\textwidth]{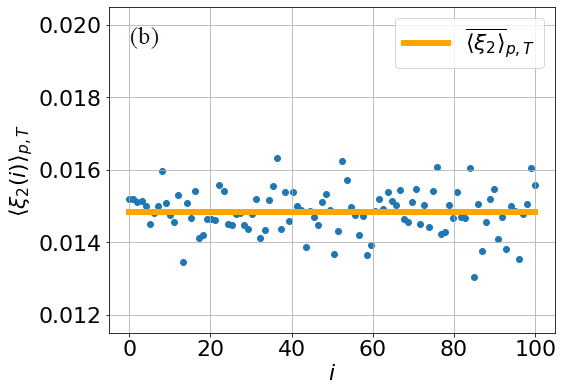} 
\includegraphics[width=0.45\textwidth]{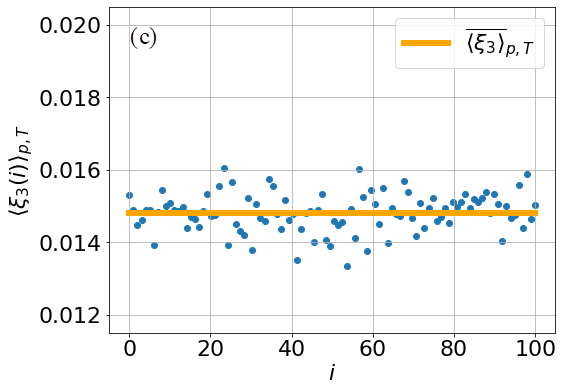} 
\caption{Reconstructions of $\xi_i(T,p)$. Deviation from the mean of the 100 individual reconstructions. Points are $\langle \xi_i(i) \rangle_{p,T}$, the solid line is $\overline{\langle \xi_i \rangle}_{p,T}$.} 
\label{calc1}
\end{figure}

Fig.~\ref{calc1} shows the deviations of the $N_{\rm rec}=100$ individual reconstructions, which are indexed with $i$, of the non-Boussinesq parameter $\langle \xi_k(i) \rangle_{p,T}$ from the mean of all calculations for $k=1, 2$ and 3. We therefore first average the reconstructed field $\xi_k(T,P)$ over the $p$--$T$ plane which is indicated by $\langle\cdot\rangle_{p,T}$. The mean non-Boussinesq parameter is eventually obtained by 
\begin{equation}
\overline{\langle \xi_k \rangle}_{p,T}=\frac{1}{N_{\rm rec}}\sum_{i=1}^{N_{\rm rec}} 
\langle \xi_k(i) \rangle_{p,T} \quad\mbox{for}\quad k=1, 2, 3\,.
\end{equation}

\end{document}